\documentclass[12pt]{article}
\usepackage{epsf}
\usepackage[utf8]{inputenc}
\usepackage[english]{babel}

\begin{document}

\title{
{\bf The hard Pomeron impact on the high-energy elastic scattering of nucleons}}
\author{A.A. Godizov\thanks{E-mail: anton.godizov@gmail.com}\\
{\small {\it SRC A.A. Logunov Institute for High Energy Physics}},\\ {\small {\it NRC ``Kurchatov Institute'', 142281 Protvino, Russia}}}
\date{}
\maketitle

\vskip-1.0cm

\begin{abstract}
The role of the hard Pomeron (HP) exchanges in the high-energy diffractive interaction of nucleons is explored. It is demonstrated that the HP subdominance at 
available energies and low transferred momenta is due to the extremely low slope of its Regge trajectory.
\end{abstract}

\section*{Introduction}

In the framework of Regge phenomenology \cite{collins}, the observed growth of the $pp$ total and elastic cross-sections at collision energies higher than 20 GeV \cite{pdg} 
is explained in terms of the soft Pomeron exchanges \cite{donnachie,donnachie2}, where the soft Pomeron (SP) is a supercritical Reggeon with the intercept of its Regge 
trajectory $\alpha_{\rm SP}(0) \approx 1.1$. By full analogy, the available data on the proton unpolarized structure function $F^p_2(x,Q^2)$ \cite{struc} at high values 
of the incoming photon virtuality $Q^2$ and low values of the Bjorken scaling variable $x$ can be described in terms of another Pomeron (called ``hard'') with the intercept 
$\alpha_{\rm HP}(0) = 1.32\pm 0.03$ \cite{godizov} or even higher (see, for instance, \cite{donnachie3}, \cite{selyugin}, or \cite{fazio}). In spite of the fact that 
$\alpha_{\rm HP}(0)>\alpha_{\rm SP}(0)$, the HP impact on the nucleon-nucleon diffractive scattering seems to be insignificant \cite{donnachie2}.

An easy way to explain this elusiveness of the HP in high-energy soft interaction is just to presume the suppression of its coupling to hadrons in the nonperturbative 
regime, which automatically leads to the SP dominance in the diffractive interaction in the absence of hard scale. However, such a physical pattern seems somewhat exotic, 
since both the Pomerons are apparently composed of gluon matter.\footnote{The leading meson Regge trajectories of the Quark Model have intercepts considerably below the 
unity. A detailed discussion of the gluon nature of supercritical Reggeons can be found in classical papers \cite{low,bfkl}.} Hence, no evident argument exists why their 
couplings to proton at low transferred momenta should differ greatly by order of magnitude, while the difference between their intercepts is large enough to expect the HP 
dominance or, at least, significance at the LHC energies. 

Below we address the problem of the HP contribution into the $pp$ high-energy elastic scattering to provide a more natural interpretation of the HP ``invisibility'' in 
the soft interaction of hadrons than the above-mentioned presumption about the coupling suppression.

\section*{The HP Regge trajectory}

First of all, let us pay attention to the behavior of the HP Regge trajectory in the asymptotic region $t\to -\infty$. 

Presumably, the HP is the leading Reggeon of the BFKL series \cite{kirschner}:
\begin{equation}
\label{BFKL}
\alpha^{(n_r)}_{\rm BFKL}(t) = 1+\frac{12\,\ln 2}{\pi}\alpha_s(\sqrt{-t})\left[1-\alpha_s^{2/3}(\sqrt{-t})\left(\frac{7\zeta(3)}{2\,\ln 2}\right)^{1/3}
\left(\frac{3/4+n_r}{11-2/3\,N_f}\right)^{2/3}+...\right]\,,
\end{equation}
where $\alpha_s(\mu)$ is the QCD running coupling, $N_f$ is the number of quark flavors, and $n_r$ is the radial quantum number.
If $t=-M_Z^2=-(91.2$ GeV$)^2$, $\alpha_s(M_Z) = 0.118$, $n_r=0$, and $N_f=$ 5 or 6, then we obtain 
\begin{equation}
\label{asymp}
\alpha_{\rm HP}(-M_Z^2)=\alpha^{(0)}_{\rm BFKL}(-M_Z^2)\approx 1.28\,.
\end{equation}
Note, that the second term in the brackets in the right-hand side of (\ref{BFKL}) is $\sim 0.1$ under the chosen values of the parameters. Thus, the estimation 
(\ref{asymp}) is quite justified.

Comparing the values of $\alpha_{\rm HP}(t)$ at $t=0$ and $t=-M_Z^2$, as well as the quantities\linebreak $\alpha'_{\rm HP}(-M_Z^2)\approx 2\cdot 10^{-6}$ GeV$^{-2}$ and 
$\frac{\alpha_{\rm HP}(0)-\alpha_{\rm HP}(-M_Z^2)}{M_Z^2}\approx 5\cdot 10^{-6}$ GeV$^{-2}$, one might come to a conclusion that both the functions $\alpha_{\rm HP}(t)$ 
and $\alpha'_{\rm HP}(t)$ evolve very slowly in the interval $-M_Z^2<t<0$. Moreover, even if $\alpha'_{\rm HP}(t)$ is essentially nonlinear in the considered range and 
$\alpha'_{\rm HP}(0)$ is, say, 100 times higher than $\alpha'_{\rm HP}(-M_Z^2)$, it is quite reasonable to consider $\alpha_{\rm HP}(t)\approx \alpha_{\rm HP}(0)$ 
at\linebreak $-3$ GeV$^2<t<0$.

Such a weak $t$-dependence is a very important feature of the HP Regge trajectory, which allows to make unambiguous conclusions on the basis of further analysis.

\section*{The HP exchange contribution into the eikonal}

\begin{table}[ht]
\begin{center}
\begin{tabular}{|l|l|}
\hline
\bf Parameter          & \bf Value         \\
\hline
$\alpha_{\rm SP}(0)-1$  & 0.109            \\
$\tau_a$                & 0.535 GeV$^2$    \\
$g_{\rm SP}(0)$         & 13.8 GeV         \\
$a_g$                   & 0.23 GeV$^{-2}$  \\
\hline
\end{tabular}
\end{center}
\vskip -0.2cm
\caption{The parameter values for (\ref{pomeron}) obtained via fitting \cite{godizov2} to the high-energy elastic scattering data.}
\label{tab1}
\end{table}

The soft-Pomeron-exchange eikonal approximation has the following structure \cite{godizov2}:
$$
\frac{d\sigma}{dt} = \frac{|T(s,t)|^2}{16\pi s^2}\,\;,\;\;\;\;T(s,t) = 4\pi s\int_0^{\infty}db^2\,J_0(b\sqrt{-t})\,\frac{e^{2i\delta(s,b)}-1}{2i}\;,
$$
\begin{equation}
\label{eikrepr}
\delta(s,b)=\frac{1}{16\pi s}\int_0^{\infty}d(-t)\,J_0(b\sqrt{-t})\,\delta_{\rm SP}(s,t) = 
\end{equation}
$$
= \frac{1}{16\pi s}\int_0^{\infty}d(-t)\,J_0(b\sqrt{-t})\;
g^2_{\rm SP}(t)\left(i+{\rm tan}\frac{\pi(\alpha_{\rm SP}(t)-1)}{2}\right)\pi\alpha'_{\rm SP}(t)\left(\frac{s}{2s_0}\right)^{\alpha_{\rm SP}(t)},
$$
where $s$ and $t$ are the Mandelstam variables, $b$ is the impact parameter, $s_0 = 1$ GeV$^2$, $\alpha_{\rm SP}(t)$ is the Regge trajectory of the soft Pomeron, 
$g_{\rm SP}(t)$ is the SP coupling to proton. At $t<0$, $\alpha_{\rm SP}(t)$ and $g_{\rm SP}(t)$ can be approximated by simple test functions 
\begin{equation}
\label{pomeron}
\alpha_{\rm SP}(t) = 1+\frac{\alpha_{\rm SP}(0)-1}{1-\frac{t}{\tau_a}}\;,\;\;\;\;g_{\rm SP}(t)=\frac{g_{\rm SP}(0)}{(1-a_gt)^2}\;,
\end{equation}
where the free parameters take on the values presented in Table \ref{tab1}. 

\begin{figure}[ht]
\epsfxsize=8.2cm\epsfysize=8.2cm\epsffile{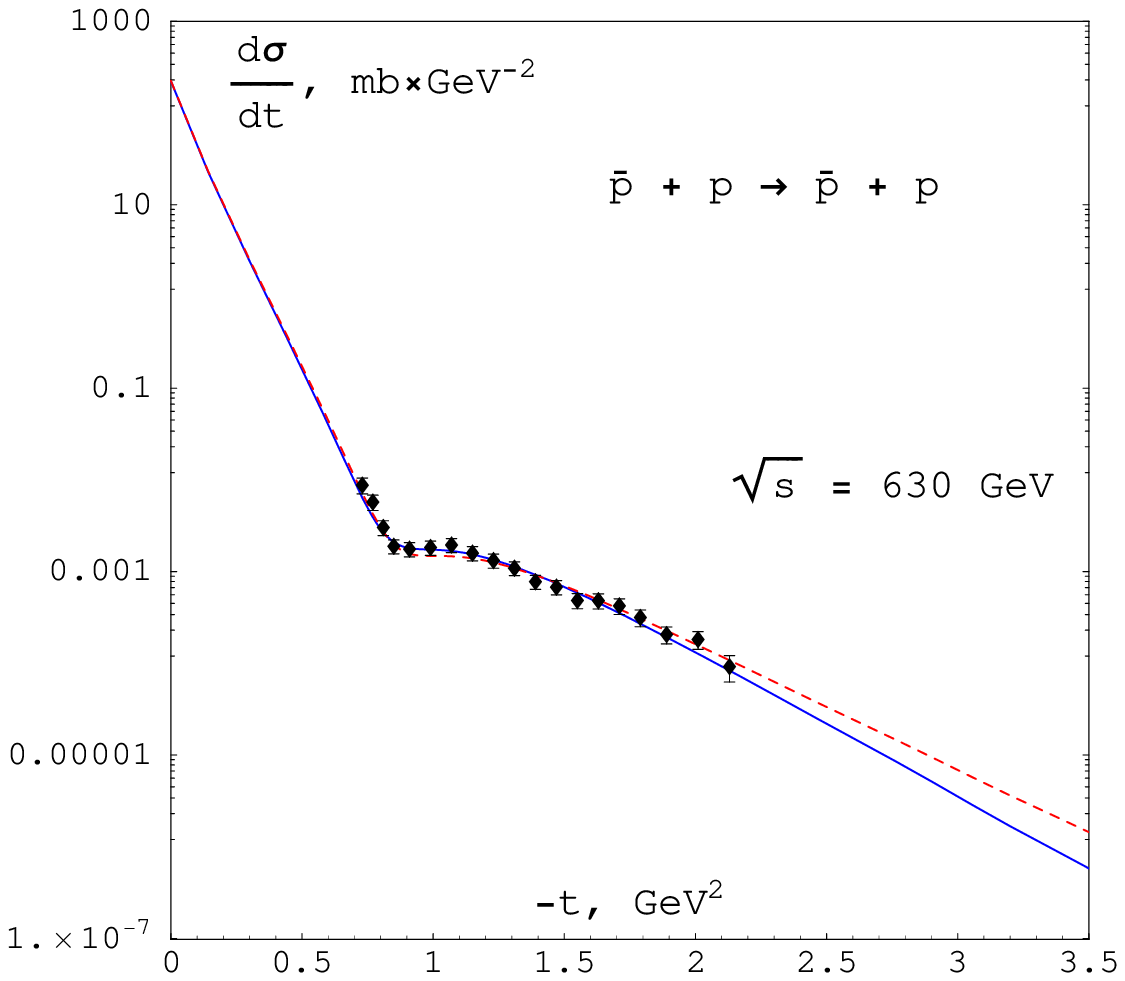}
\vskip -8.2cm
\hskip 8.5cm
\epsfxsize=8.15cm\epsfysize=8.15cm\epsffile{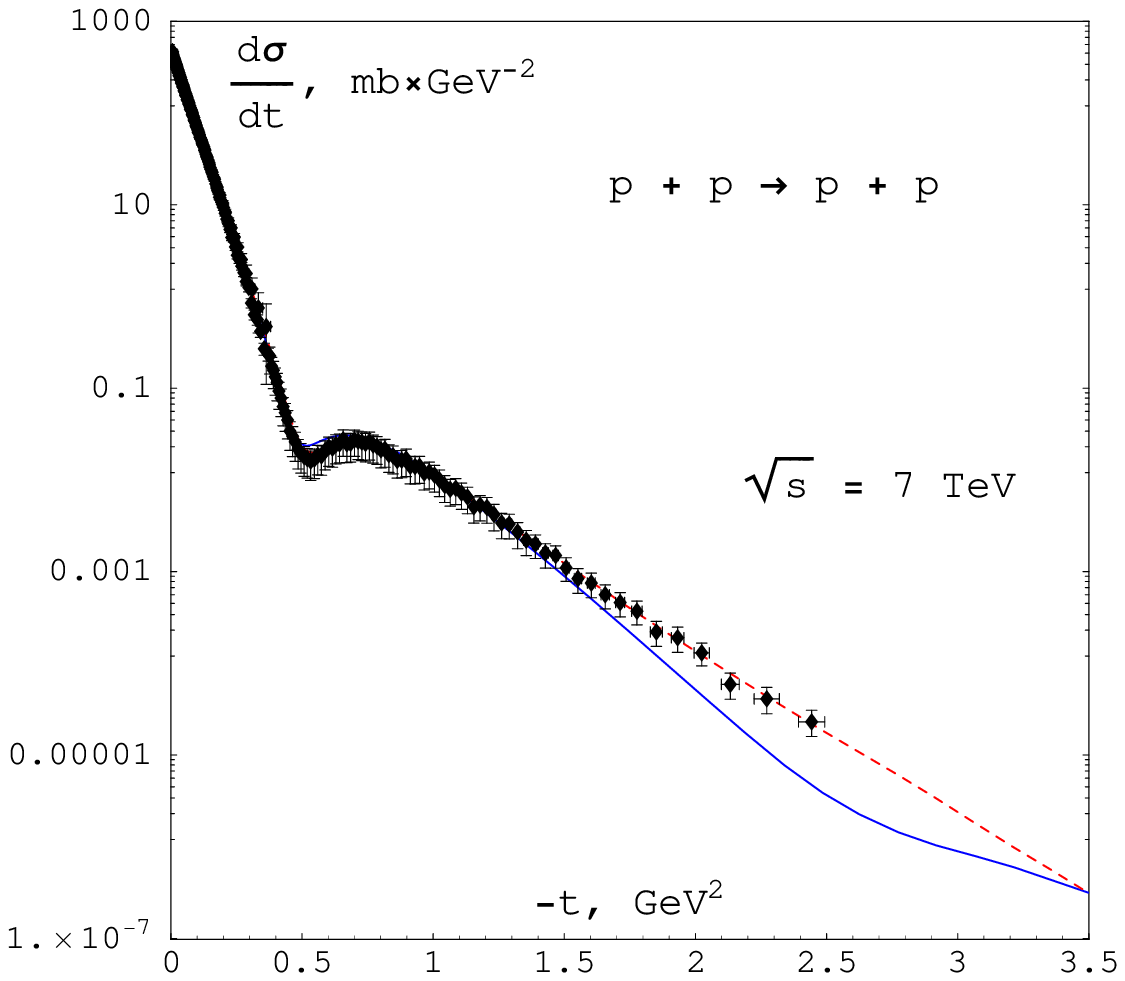}
\vskip -0.2cm
\epsfxsize=8.2cm\epsfysize=8.2cm\epsffile{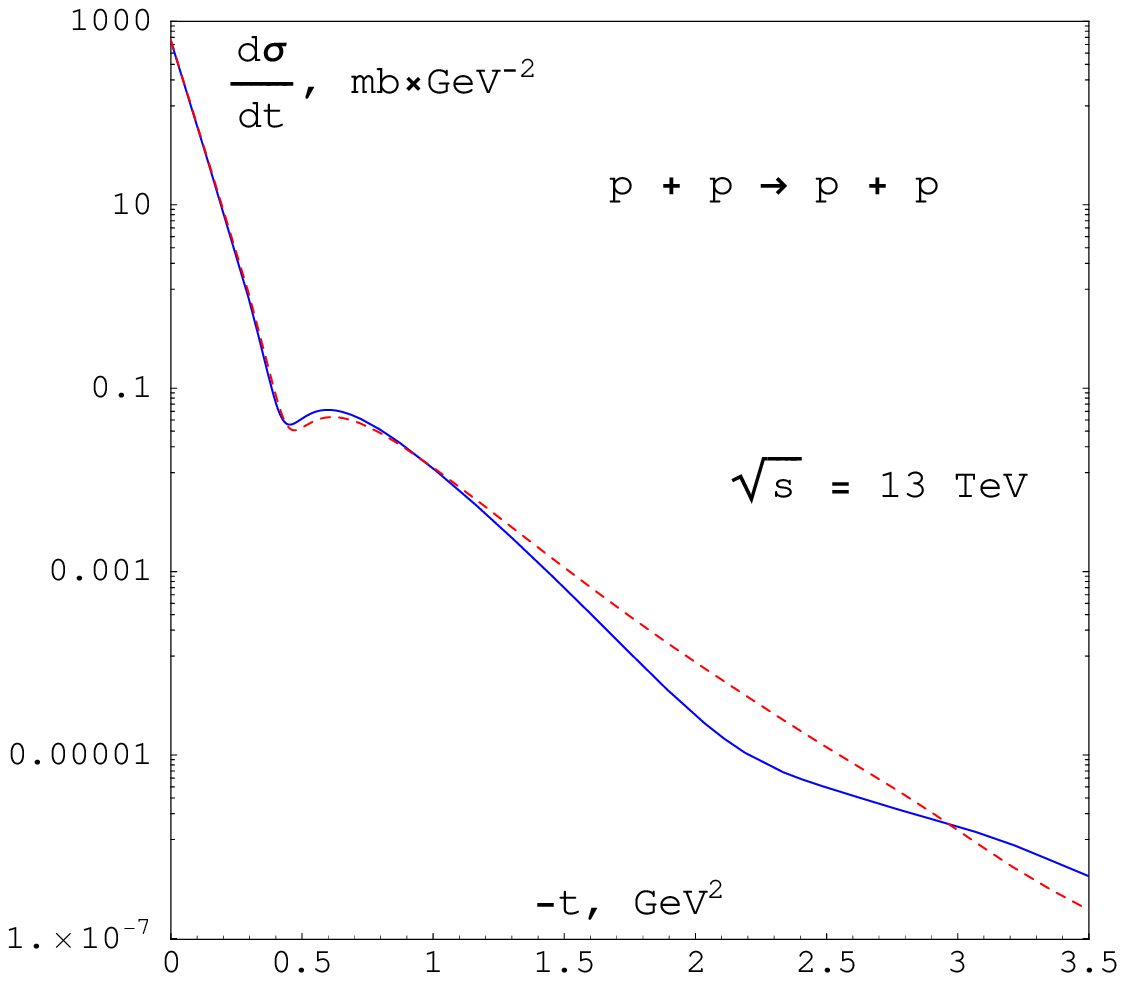}
\vskip -7.95cm
\hskip 9.1cm
\epsfxsize=7.9cm\epsfysize=7.9cm\epsffile{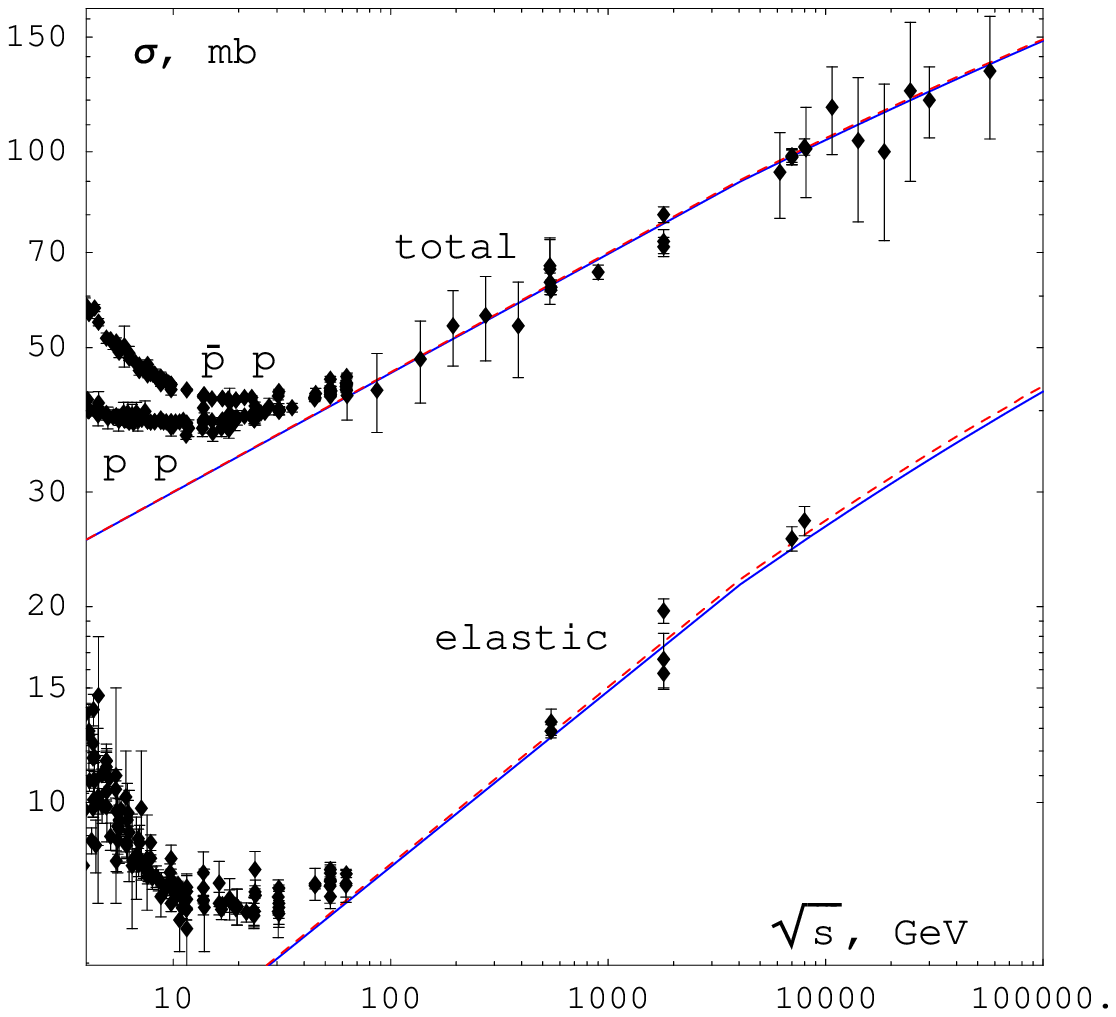}
\vskip -0.5cm
\caption{The HP impact on the observables of nucleon-nucleon scattering \cite{pdg,ua4,totatl}. The used HP parameters are $\alpha_{\rm HP}(0)=1.32$, 
$\beta_{\rm HP}(0)=0.08$, $b=1.5$ GeV$^{-2}$. The solid (dashed) lines correspond to the model ignoring 
(taking account of) the HP exchanges.}
\label{diff}
\end{figure}

Inclusion of the HP exchanges into consideration requires a replacement\linebreak $\delta_{\rm SP}(s,t)\to \delta_{\rm SP}(s,t)+\delta_{\rm HP}(s,t)$, where
\begin{equation}
\label{eikhard}
\delta_{\rm HP}(s,t)=\left(i+{\rm tan}\frac{\pi(\alpha_{\rm HP}(0)-1)}{2}\right)\beta_{\rm HP}(t)\left(\frac{s}{2s_0}\right)^{\alpha_{\rm HP}(0)}\,.
\end{equation}
Choosing $\alpha_{\rm HP}(0)=1.32$ and $\beta_{\rm HP}(t)=\beta_{\rm HP}(0)\,e^{b\,t}$, where $\beta_{\rm HP}(0)=0.08$ and $b=1.5$ GeV$^{-2}$, we come to the pattern 
presented in Fig. \ref{diff}. The description quality is satisfactory: for example, $\Delta \chi^2\approx 12$ over 19 points of the data set \cite{ua4} and 
$\Delta \chi^2\approx 215$ over 205 points of the data set \cite{totatl}. The description of other data considered in \cite{godizov2} remains satisfactory 
as well.

As we see, the account of the HP exchanges improves the description of $d\sigma/dt$ at\linebreak $\sqrt{s}=$ 7 TeV without any refitting of the SP parameters. Regarding 
available data at the lower energies, the HP impact can be ignored though.

\section*{Discussion and conclusions} 

The HP subdominance at accessible energies is, certainly, determined by the smallness of its Regge residue: 
\begin{equation}
\label{hardres}
\beta_{\rm HP}(t) = g^2_{\rm HP}(t)\,\pi\alpha'_{\rm HP}(t)\,.
\end{equation}
Assuming that $4<\alpha'_{\rm HP}(0)/\alpha'_{\rm HP}(-M_Z^2)<100$, we obtain $g_{\rm HP}(0)\sim g_{\rm SP}(0)$, what is quite natural in view of the presumed 
glueball nature of both the Pomerons. The smallness of $\beta_{\rm HP}(t)$ at low negative $t$ is, thus, related to the extremely weak $t$-behavior of $\alpha_{\rm HP}(t)$.

The low $t$-slope of $\alpha_{\rm HP}(t)$ may take place in the region $t>0$ as well. It would imply the existence of some series of ultraheavy resonances lying on the 
HP Regge trajectory. Due to their spin properties, such an ultraheaviness (tens or hundreds of GeV) accomplished by strong enough coupling to light hadrons inevitably 
results in the ultrashort life of the HP resonance states. The conception of heavy Pomeron is, certainly, not new. It was proposed 
by\linebreak V.N. Gribov more than 40 years ago \cite{gribov}. The only difference between Gribov's heavy Pomeron and the BFKL HP is in their intercept values. 

Above, we neglected the impact of the subleading (daughter) Pomerons corresponding to nonzero values of $n_r$ in series (\ref{BFKL}). The reason is that 
the leading Pomeron intercept is separated from the subleading ones by significant gap \cite{heckathorn}. A similar pattern takes place for other known series of Reggeons 
in asymptotically free field theories \cite{lovelace,godizov3}. Moreover, as the Regge trajectories are expected to be Herglotz functions \cite{collins}, so the 
contributions of the subleading BFKL Pomerons at nonzero $n_r$ and low negative $t$ are suppressed in the factors $\alpha^{(n_r)'}_{\rm BFKL}(t)$ (as compared to 
$\alpha^{(0)'}_{\rm BFKL}(t)$) in addition to the suppression in the values of $\alpha^{(n_r)}_{\rm BFKL}(t)$. The much higher slope of $\alpha_{\rm SP}(t)$ points to the 
fact that the soft Pomeron is not a Reggeon from the BFKL series.

In view of the aforesaid, we come to the main conclusion:
\begin{itemize}
\item The conception of the hard Pomeron as the leading Reggeon of the BFKL series is quite consistent with the available data on the high-energy $pp$ elastic scattering. 
Its ``invisibility'' at the collision energies lower than 2 TeV is related not to the smallness of its coupling to proton (which is of the same order as the soft Pomeron's 
one) but to its extremely weak $t$-evolution in the scattering region. In its turn, such a weak $t$-behavior seems to be related to a possible ultraheaviness of the 
resonances corresponding to this Reggeon. Hence, the characteristics ``light'' and ``heavy'', regarding these two Pomerons, seem to be more natural than ``soft'' and 
``hard'' (though, it is just a matter of conventions in terminology).
\end{itemize}

\begin{figure}[ht]
\epsfxsize=8.2cm\epsfysize=8.2cm\epsffile{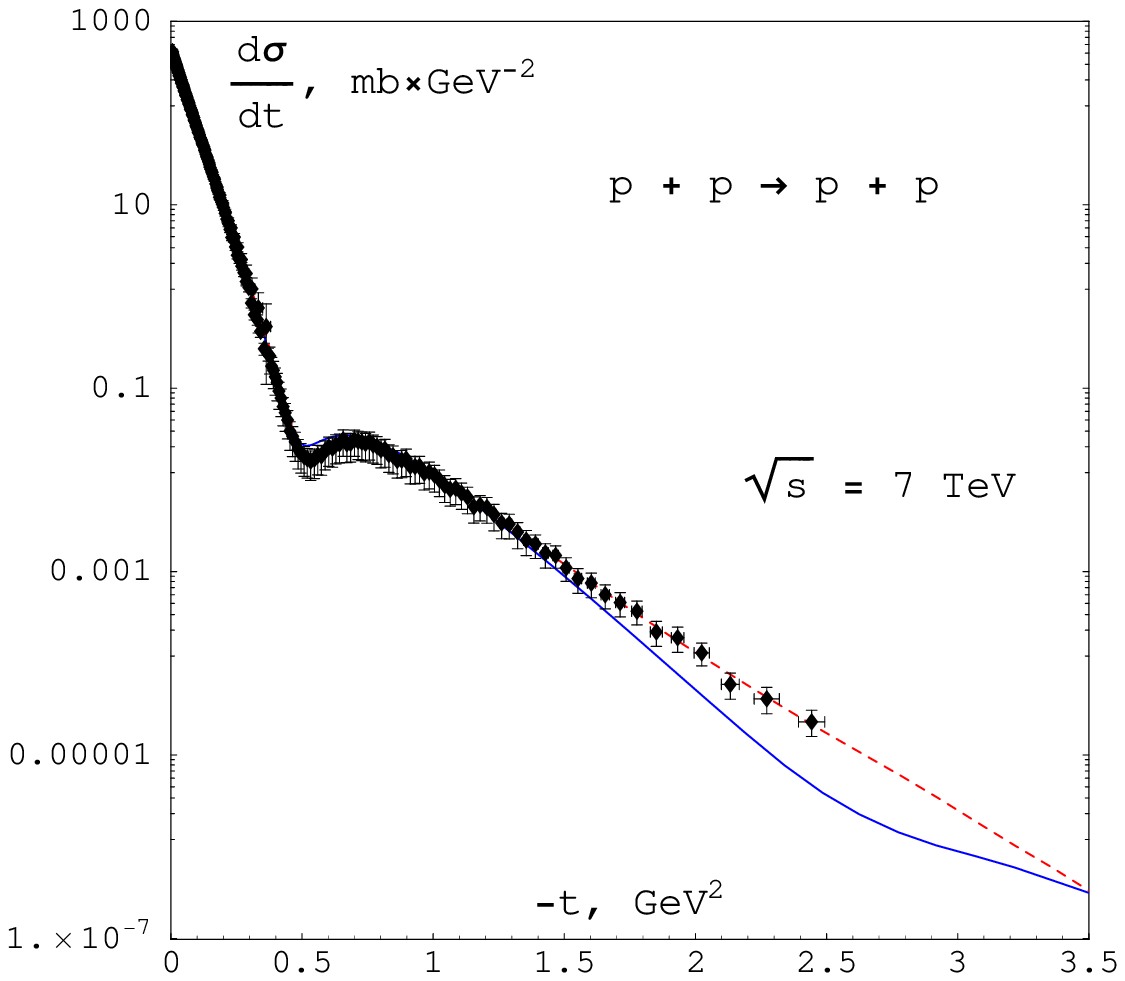}
\vskip -8.2cm
\hskip 8.5cm
\epsfxsize=8.15cm\epsfysize=8.15cm\epsffile{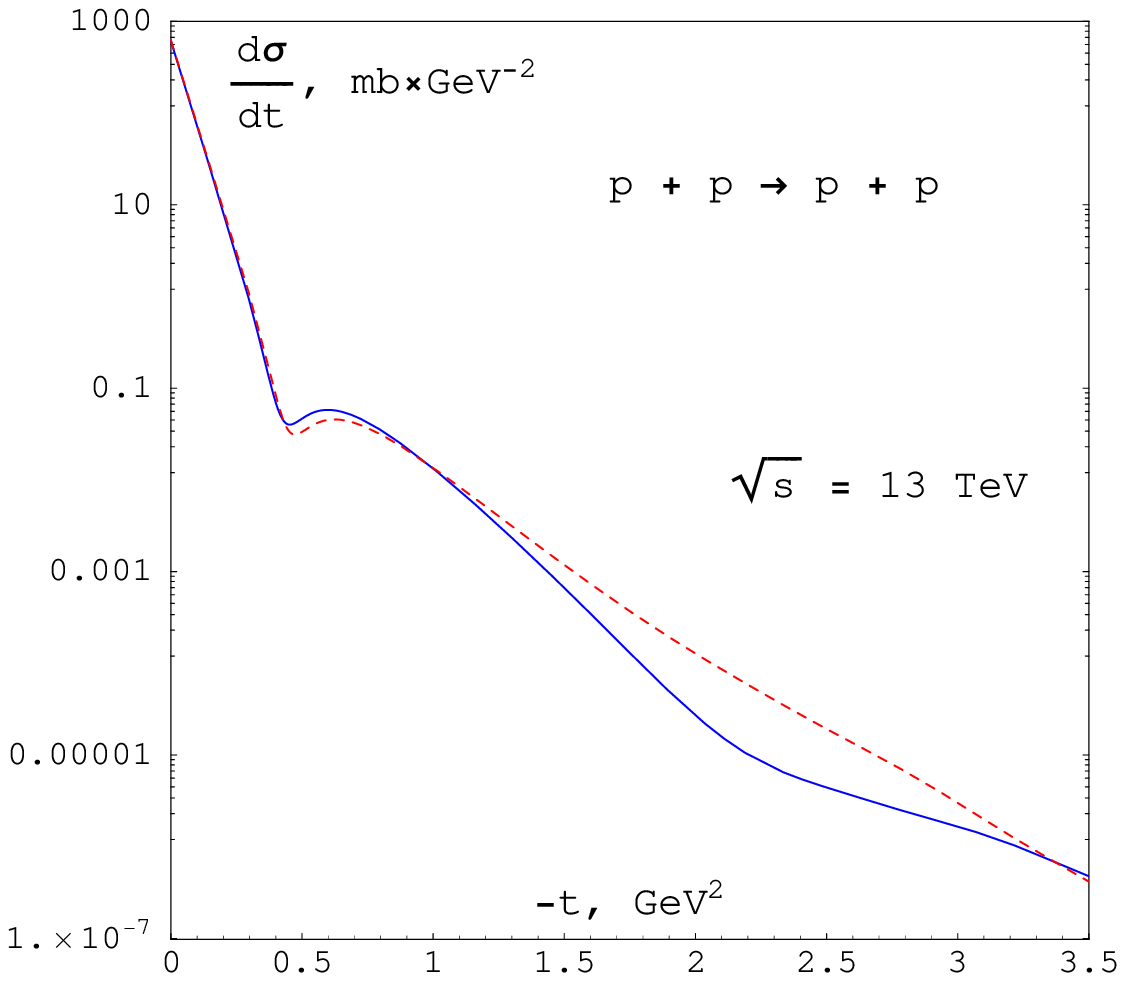}
\caption{The HP impact on the differential cross-section of $pp$ elastic scattering at the LHC energies. The used HP parameters are $\alpha_{\rm HP}(0)=1.44$, 
$\beta_{\rm HP}(0)=0.01$, $b=1.5$ GeV$^{-2}$. The solid (dashed) lines correspond to the model ignoring 
(taking account of) the HP exchanges.}
\label{diff2}
\end{figure}

In the very end, it should be pointed out that the TOTEM data at $\sqrt{s}=$ 7 TeV only do not allow to confirm or discriminate the used phenomenological estimation 
of the HP intercept. The problem of absorption was swept under the carpet in \cite{godizov}, though the relative contribution of the absorptive corrections may be 
non-vanishing in the kinematic range considered in \cite{godizov}. Therefore, the true value of $\alpha_{\rm HP}(0)$ may be a bit higher and, so, the estimation 
$\alpha_{\rm HP}(0) = 1.32\pm 0.03$\linebreak  should be treated just as the lower bound for this quantity. For example, the variant\linebreak $\alpha_{\rm HP}(0)=1.44$ 
\cite{donnachie3} and $\beta_{\rm HP}(t)=\beta_{\rm HP}(0)\,e^{b\,t}$, where $\beta_{\rm HP}(0)=0.01$ and $b=1.5$ GeV$^{-2}$, also yields a satisfactory description of the 
TOTEM data (see Fig. \ref{diff2}). The data at $\sqrt{s}=$ 13 TeV are needed for more or less reliable determination of $\alpha_{\rm HP}(0)$.

In any case, the account of the HP exchanges extends the applicability range of the Regge-eikonal approximation (\ref{eikrepr}),(\ref{pomeron}) for the elastic scattering 
of nucleons at ultrahigh energies. The satisfactory reproduction of available data by the updated model demonstrates the incorrectness of the claim \cite{troshin} 
that absorptive models do not provide a good description of the LHC data in the deep-elastic scattering region.

\section*{Acknowledgments} The author thanks V.A. Petrov and R.A. Ryutin for discussions.

\end{document}